\documentclass[aip,rsi,reprint,graphicx]{revtex4-1} 
\usepackage{graphicx}

\draft 

\begin{document}


\title{Cryogenic infrared filter made of alumina for use at millimeter wavelength} 



\author{Yuki Inoue}
\email[]{iyuki@post.kek.jp}
\affiliation{The Graduate University for Advanced Studies, Ibaraki 305-0801, Japan}
\author{Tomotake Matsumura}
\affiliation{High Energy Accelerator Research Organization, Ibaraki 305-0801, Japan}
\author{Masashi Hazumi}
\affiliation{The Graduate University for Advanced Studies, Ibaraki 305-0801, Japan}
\affiliation{High Energy Accelerator Research Organization, Ibaraki 305-0801, Japan}
\author{Adrian T. Lee}
\affiliation{University of California, Berkeley, Berkeley CA 94720, USA}
\author{Takahiro Okamura}
\affiliation{High Energy Accelerator Research Organization, Ibaraki 305-0801, Japan}
\author{Aritoki Suzuki}
\affiliation{University of California, Berkeley, Berkeley CA 94720, USA}
\author{Takayuki Tomaru}
\affiliation{High Energy Accelerator Research Organization, Ibaraki 305-0801, Japan}
\author{Hiroshi Yamaguchi}
\affiliation{High Energy Accelerator Research Organization, Ibaraki 305-0801, Japan}

\date{\today}
\begin{abstract}
We propose a high thermal conductivity infrared (IR) filter using alumina for use in millimeter wave detection systems.
We constructed a prototype two-layer anti-reflection (AR) coated alumina filter with a diameter of 100~mm and a thickness of 2~mm, and characterized its thermal and optical properties.
The transmittance of this filter at 95~GHz and 150~GHz is 97~\% and 95~\% while the estimated 3~dB cutoff frequency is at 450~GHz. 
The high thermal conductivity of alumina minimizes thermal gradients.
We measure a differential temperature of only 0.21~K between the center and the edge of the filter when it is mounted on a thermal anchor of 77~K. We also constructed a thermal model based on the prototype filter and analyzed the scalability of the filter diameter. We conclude that temperature increase at the center of alumina IR filter is less than 6~K even with a large diameter of 500~mm, when the temperature at the edge of the filter is 50~K. This is suitable for an application to a large-throughput next-generation cosmic microwave background (CMB) polarization experiment, such as POLARBEAR-2 (PB-2).
\end{abstract}

\keywords{Cosmic Microwave Background, Millimeter wave, Gravitational Waves, IR filter, AR coating, POLARBEAR-2}

\maketitle 

\section{Introduction}
Measurements of the cosmic microwave background (CMB) polarization have been providing essential information for studying the early universe~\cite{Kermish:2012eh}. The odd parity pattern imprinted in the CMB polarization, called the B-mode, from the last scattering surface is particularly important as a tool to probe a signal from cosmic inflation. The community-wide effort to hunt this B-mode is in progress. The expected signal fluctuations are at the nK level, which is much smaller than the CMB black body radiation temperature (2.7~K).
 Thus an instrument to measure this tiny signal should have high statistical sensitivity with stringent control of systematics over broad spectral coverage. 

One path to realize such an instrument is to collect more photons, i.e. increasing the throughput of a telescope with cryogenically cooled superconducting detector arrays~\cite{Hanany:2012vj}. The sensitivity of a cutting-edge superconducting detector is limited by the background photons, and the key to increase the sensitivity is to increase the number of detectors. Next-generation experiments will have an array of $\sim 10^4$ detectors at a sub-Kelvin stage. 
While this is straightforward conceptually, the actual implementation comes with technical challenges. 
A receiver with cooled optical system tends to have a large window to receive incoming photons. Although it is ideal that only the CMB photons at millimeter wave pass through, the radiation from other frequencies, including infrared (IR), can also pass through and it contributes as a significant heat load~\cite{2006SPIE.6275E..25T}. 

The typical solution for this problem is to use an absorptive IR filter that is made of plastic, such as PTFE, or Nylon~\cite{Gonz,2010SPIE.7741E..40O, 2010SPIE.7741E..50S,Filippini:2011ds, Hileman:2011}. 
These filters absorb the incident IR radiation and the absorbed heat is conducted through the filter to the edge where the filter is mounted. 
The temperature of the filter is determined by the thermal conductance of the filter and the absorption of the incident radiative power. A filter made of plastic has a poor thermal conductivity and the temperature of the filter becomes higher as the filter diameter increases. 
As a result the filter itself becomes the dominant emissive element in the optical path. This results in both/either exceeding the available cooling power of a cryostat and/or degrading the detector sensitivity due to the higher optical loading. Another standard solution is to place a reflective IR filter, such as metal-mesh filter~\cite{2006SPIE.6275E..25T}. It is yet limited in an available diameter above $\sim500$~mm due to the fabrication difficulty. 

In order to solve the issues of thermal conductivity and the available size for a large throughput receiver, we propose a new IR filter made of alumina. Alumina has ideal properties: i) low optical loss at millimeter wavelength, ii) absorbing IR radiation efficiently; iii) high thermal conductivity at 100~K; and iv) availability of large-diameter plates.
We describe the design of the prototype IR filter using alumina and its thermal and optical characterizations. We also discuss the impact on the power to the detector and the cooling power when this filter is used for a next generation CMB polarization experiment, POLARBEAR-2 (PB-2)~\cite{Tomaru:2012,Matsumura:2012}.

\section{Design goals and filter design}
\subsection{Design goals}
We developed a prototype alumina filter with a diameter of 100~mm and characterized its thermal and optical properties. 
We establish the design goals that meet requirements in the next-generation CMB experiments as following.
\begin{itemize}
	\item 95~\% transmittance at the detection bands, i.e. 95 and 150~GHz with $30~\%$ fractional bandwidth,
	\item 3~dB cutoff frequency below 1~THz,
	\item the temperature of the filter less than 200~K with the bath temperature of 50~K,
	\item extendability of the filter diameter to 500~mm or larger.
\end{itemize}
As an example, one of the next-generation experiments, PB-2, requires a window diameter of 490~mm and the corresponding required IR filter diameter is about 500~mm. We discuss the extendability in more detail in Section~\ref{Disc}.

\subsection{Alumina properties}
We use samples of sintered polycrystalline alumina manufactured by Nihon Ceratech with purity of 99.5~\% (as they label 99.5LD)~\cite{Nihon}. The disk-shaped alumina sample has the diameter of 100~mm and thickness of 2~mm. Basic thermal and optical properties of the alumina sample, which are adequate to describe this work, are summarized in Table \ref{fig:thermal_conductivity} and \ref{material}. More detailed descriptions on the alumina properties are given elsewhere~\cite{Material_YI_TM}.

\begin{table}
\caption{The measured thermal conductivity of alumina as a function of temperature. The thermal conductivity of Nylon, PTFE and quartz are also listed for comparison. The thermal conductivity of alumina is three orders of magnitude as high as that of plastic, such as PTFE and Nylon which is commonly used for millimeter wave. This is also a factor of 4 higher than that of quartz~\cite{NIST,simon:1994}. \label{fig:thermal_conductivity}}
 \begin{tabular}{c|c|c} \hline
    Material & Temperature & Thermal conductivity \\ 
    &[K]& [$\mathrm{W/m \cdot K}$] \\ \hline
    Alumina &77-90 &  $144 \pm 35$  \\
    PTFE & 77 & 0.13 \\
    Nylon & 77 & 0.29 \\ 
    Quartz & 59.2 & 18.7 \\ \hline
  \end{tabular}
\end{table}

 \begin{table} 
  \caption{ Basic properties of our alumina filter sample. The thickness, d, the reflective index, $n$, the loss tangent, tan~$\delta$, and the coefficient of thermal expansion, $\Delta L/L$, are listed. All values are from our measurements, except for $\Delta L/ L$ of Stycast 2850FT and 1090 which are listed from the technical data sheet of Emerson and Cuming~\cite{EandC}. Our measurements for $n$, $\tan{\delta}$ and $\Delta L/L$ are at the liquid nitrogen temperature,
while the thickness values are at the room temperature. The optical properties of alumina are measured by using a variable millimeter-wave source between 72 and 110 GHz. We derived the index and loss tangent from the transmittance as a function of the wavelength as described in Inoue et al \cite{Material_YI_TM}. The errors include both statistical and systematic uncertainties. The systematic errors arise from the uncertainty of the AR thicknesses and the fluctuations of the measurement system due to the temperature variation.  \label{material}} 
 \begin{tabular}{c|c|c|c|c} \hline
    Material & $d$ & $n$ & $\tan{\delta}$ & $\Delta L/L$ \\ 
    &[mm]&& [$\times 10^{-4}$] & [\%]\\ \hline
    Alumina & $2.05^{\pm0.005}$ & $ 3.117^{\pm0.005}$ & $3.0^{\pm1.1}$  & $0.049^{\pm 0.002}$ \\
    Stycast 2850FT & $0.27^{\pm0.015}$ & $ 2.196^{\pm 0.004}$  & $33^{\pm3}$  &  0.3 \\
    Stycast 1090 & $0.42^{\pm0.015}$ & $ 1.423^{\pm 0.025}$  &  $57^{\pm8}$ &  0.2-0.3 \\ \hline
  \end{tabular}
\end{table}

\subsection{Filter fabrication}
\begin{figure}
\includegraphics[width=6.6cm]{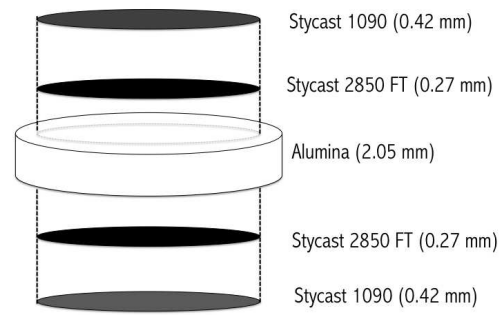}
\caption{A schematic view of the alumina IR filter configuration. The sample diameter is 100~mm and the AR layers are extended to the diameter of 90~mm. The surface of alumina at the filter edge of 5~mm is without the IR layers, and used as the mount region in order to cool alumina filter directly.\label{fig:af_config}}
\end{figure}

Figure~\ref{fig:af_config} shows the alumina filter configuration. Our design consists of alumina that has two-layer anti-reflection (AR) coating on both sides. 
 The AR coating is designed to maximize the transmittance of the normal incident radiation at two detection bands, 95 and 150~GHz, each with 30~\% bandwidth. The two-layer AR coating consists of the epoxy glues, Stycast 2850 FT and Stycast 1090, manufactured by Emerson and Cuming~\cite{Suzuki:2012bf,Rosen:2013zza, EandC}. We measure their indices and the losses as shown in Table~\ref{material}. 
 A few thin grooves are machined on the first layer of the AR coating in order to release the mechanical stress from the differential thermal contraction among the AR layers and alumina at the cryogenic temperature. We thermally cycled the alumina filter between cryogenic temperatures and room temperature ten times. The filter survived without change in its performance of transmittance and polarization.

\section{Experiments and Results}
We perform three experiments to characterize the alumina filter. Each experiment is designed to address the following item: i) transmittance and emissivity at the detection bands; ii) absorption in the IR wavelength; iii) the integrated transmittance over all the frequencies and the filter temperature.

\subsection{Transmittance and emissivity at millimeter wavelength}

\begin{figure}
\includegraphics[width=9cm]{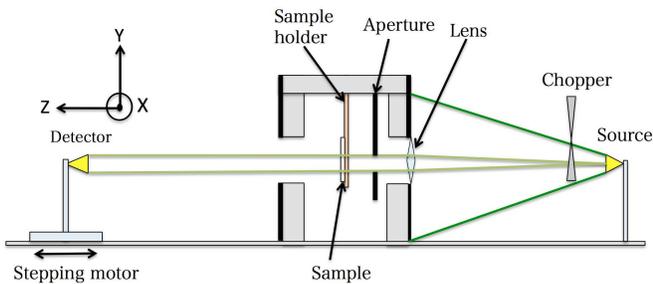}
\caption{A schematic view of the transmittance measurement system.The beam is collimated with a lens and incident to the sample, where the lens material is the AR coated rexolite and its diameter is 50~mm. The  stepping motor scans along the $z$-direction to measure and subtract the effect of the standing wave.\label{fig:OPT}}
\end{figure}

\begin{figure}
\includegraphics[width=9.5cm]{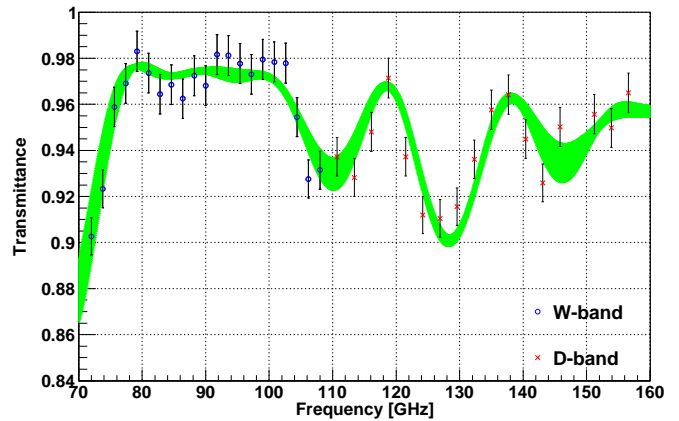}
\caption{The transmittances of the IR filter are shown as a function of the frequency. The sample is cooled at a temperature of 81~K. The green band is the prediction with 1-$\sigma$ error, which includes the errors in the construction parameters given in Table~\ref{material}.  \label{fig:AR}}
\end{figure}

Figure~\ref{fig:OPT} shows the schematic view of the transmission measurement system. We use a millimeter wave source that is tunable in frequency. A synthesized frequency generator with a sixfold or ninefold frequency multiplier covers the W- and D-bands~\cite{AMT}. The millimeter wave source is chopped, and the amplitude of the modulated signal is detected by a diode detector. This signal is demodulated  by a lock-in amplifier. A copper sample holder, which is cooled with liquid nitrogen, clamps the alumina filter and cools it to 81~K by conduction. The same cooling method is used elsewhere~\cite{Rosen:2013zza}. At each incident frequency, we measure the output voltage with and without the sample and obtain the transmittance as their ratio.
Figure~\ref{fig:AR} shows the results.
The green values are prediction, which is obtained from the parameters and their errors listed in Table~\ref{material}~\cite{Hecht}.  The predicted curve was derived using characteristic matrix method with measured index of refraction and loss-tangent \cite{OSH}. The measured transmittances within the 95~GHz and 150~GHz bands are $96.5\pm0.2~\%$ and  $94.6\pm 0.3~\%$, respectively.

We characterize the emissivity using transmittance and estimated reflectance. The estimated reflectances are 1~\% and 2~\% at 95~GHz and 150~GHz bands. These are computed from the measured construction parameters in Table~\ref{material}. 
The emissivity, $\epsilon$, is estimated by the following equation: 
\begin{equation}
	\epsilon= A =1-R( d, n, \tan{\delta}, \nu)-T( d, n, \tan{\delta}, \nu),
\end{equation}
where $A$, $R$ and $T$ are the absorption, the power reflectance and transmittance of the filter. 
The estimated emissivities are 2.5~\% and 3.5~\% in 95~GHz and 150~GHz bands, respectively.

\subsection{Transmittance at THz wavelength}
We measure the transmittance of the alumina filter at THz wavelengths using a Martin-Puplett Fourier transform spectrometer (FTS).
A mercury lamp is used as a thermal source, and an InSb bolometer is used as a detector. The detailed description of this FTS is found elsewhere~\cite{1998PASJ...50..359M}. We measure the transmittance from 250 GHz to 1500 GHz at two different temperatures of 30 K and 300 K. We prepare the alumina filters that have a thickness of 2 mm with diameter of 20 mm and 50 mm at 30 K and 300 K, respectively. It has one layer of AR coating using Stycast 2850FT ($d=0.26$~mm) on both sides.  
 
\begin{figure}
\includegraphics[width=9.5cm]{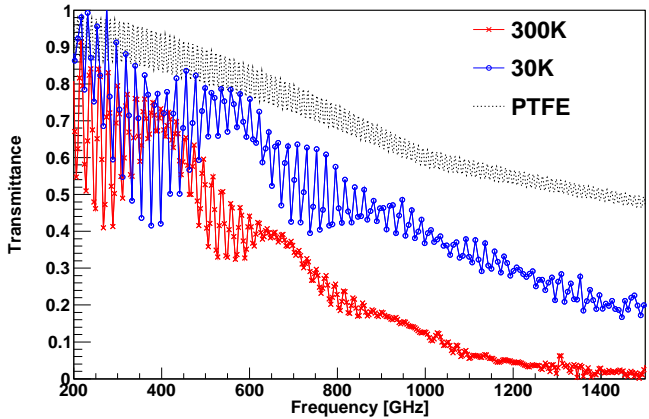}
\caption{The transmittance of the IR filter at the THz band. The crosses are for the sample temperature of 300~K and the open circles are for 30~K. This sample only has one layer of AR coating, Stycast 2850. The dashed curve is the calculated transmittance of a PTFE filter.\label{fig:IR} }
\end{figure}

Figure~\ref{fig:IR} shows the measured spectra. The data show that the spectrum of the 30~K sample has higher transmittance than the spectrum at 300~K. This is because the loss tangent of the filter, i.e. the loss from alumina and AR coating material, becomes lower at the lower temperature. 

The corresponding 3~dB cutoff frequencies of the filter are 450 and 700~GHz at 300~K and 30~K, respectively. These 3~dB cut off frequencies are sufficiently lower than our design goal. For the actual implementation, the 3~dB cutoff frequency of the two-layer AR coated alumina filter is expected lower at 30~K due to the additional loss from another layer.  The estimated 3~dB cutoff frequency is 450~GHz for the two-layer AR coating, where we have used $3.4 \times 10^{-3}$ as the loss tangent of Stycast 1090. 

For comparison, we also show the transmittance of PTFE in Fig~\ref{fig:IR} computed with the assumptions of the 20~mm thickness and the optical properties taken from elsewhere~\cite{Lamb1996c} ,where the loss tangent of PTFE value at 300~K is used. It is seen that the alumina filter cuts off the THz waves more than the PTFE filter.

\subsection{Thermal gradient of filter}

\begin{figure}
\includegraphics[width=9cm]{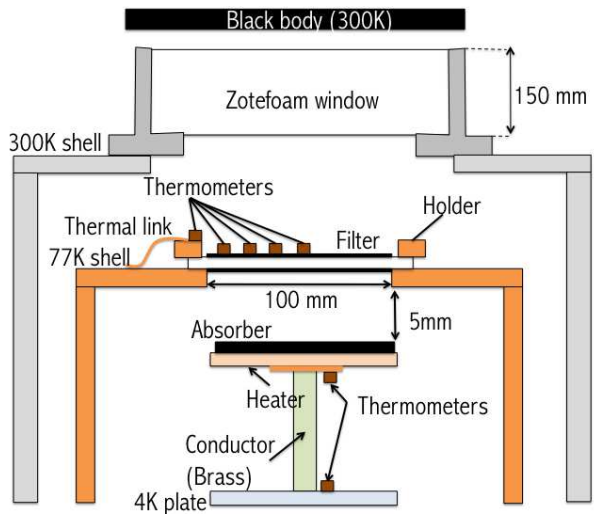}
\caption{A schematic view of experimental setup for measurements of the radial temperature profile of the filter and the integrated transmitted power to the 4~K stage. The alumina IR filter is mounted on the 77~K shell. \label{fig:thermo_config} }
\end{figure}

We measure the temperature of the prototype filter when it is used as an IR filter in a cryostat. Figure~\ref{fig:thermo_config} shows the measurement configuration. The prototype filter is mounted on a 77~K shell that is located just below the Zotefoam window. We mount the four silicon diode thermometers on the filter as shown in Figure~\ref{fig:thermo_config}. The size of thermometers is 1mm x 2mm x 3mm. The absorption of the filter is small and the  temperature of the thermometer is only depends on the filter temperature. 

\begin{figure}
\includegraphics[width=9.5cm]{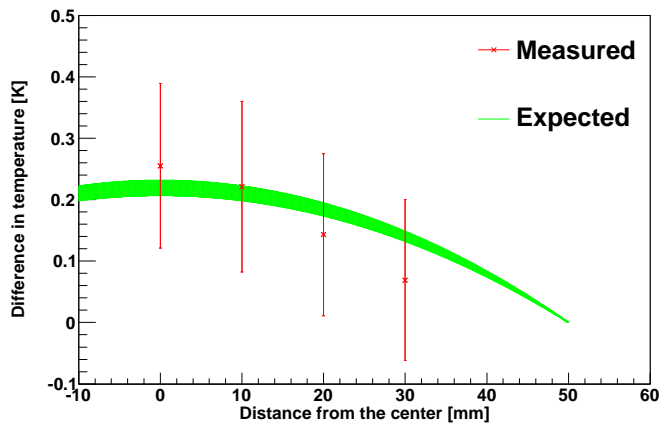}
\caption{The temperature distribution across the 100~mm alumina filter is plotted as a function of the radial distance from the center of the filter. The offset of the temperature is removed and the offset is the edge temperature of 90.3~K. The crossed points are the measured values with the error bars estimated from the calibration accuracy. The expected region is also shown in the plot, which is not a fit but is based on the thermal model with the basic parameters, in Table~\ref{fig:thermal_conductivity} and \ref{material}. \label{fig:thermal}}
\end{figure}

Figure~\ref{fig:thermal} shows the radial temperature profile of the filter. The measured temperature difference between the center and the edge is $0.26 \pm 0.13~\mathrm{K}$, where the error is from calibration accuracy of the thermometers. 

We model this temperature profile in order to make a comparison between the measured results and the expected temperature distribution that is based on the thermal conductivity. The difference of the temperature between the center and the edge can be written analytically as  
\begin{equation}
\Delta T(r)=T(r)-T_{edge}=P_{con} \frac{R^2 -r^2}{4\pi t \kappa R^2},
\label{eq:deltaT}
\end{equation}
where $T_{edge}$ is the edge temperature, $t$ is the filter thickness, and $\kappa$ is the thermal conductivity of alumina, $R$ and $r$ are the radius of the filter and the radial distance from the center~\cite{Heat_transfer}. While the filter has two-layers of AR coating we neglect them in our thermal model because they do not contribute due to the significantly lower thermal conductivities and thicknesses than those of alumina. The conducted power, $P_{con}$, from the filter to the thermal bath is assumed as 
 \begin{eqnarray}
P_{con} = P_{in}-P_{rad} \sim P_{in}
\label{eq:Peq}
\end{eqnarray}
where $P_{in}$ is the incident absorbed radiative power from the window, and $P_{rad}$ is the emissive power from the filter surface.
We use the incident absorbed radiative power from the window instead of one from the outside of the cryostat. This is because our window material, Zotefoam, is highly absorptive and it absorbs the IR radiation. The dominant source of IR radiation to the inside of the cryostat is the thermal emission of the window, which is mounted at the room temperature. The surface that is facing the inside of the cryostat is radiatively cooled. The measured temperature at the center of the surface of the window is 240~K. The expected radiative power from the window is 1.1~W. This is at least an order of magnitude higher than the power radiated from the filter with its expected temperature range, and thus $P_{rad}$ is negligible.
The measured filter temperature at the center is 90.5~K. The predicted temperature difference between the center and the edge is 0.21~K, which is consistent with the measured value as shown in Figure~\ref{fig:thermal}. 

We place an absorber that is made of Stycast 2850 and charcoal as shown in Figure~\ref{fig:thermo_config}. The detailed recipe of this absorber is found elsewhere~\cite{Bock:1994}. The reflection is expected to be 5~\% at 1~THz and lower for higher frequency. Our absorber is roughened on its surface and the incident flux is peaked above 1~THz. Thus, we use 5~\% as the conservative estimate of the reflectance for our absorber.
The absorber is connected with the brass conductor whose thermal conductivity is $4~\mathrm{W/m \cdot K}$ around 4~K~\cite{NIST}. The length and diameter of brass are 140~mm and 10~mm, and the thermal conductance is $0.002~\mathrm{W/K}$.
By measuring temperature difference between the absorber and the 4K plate, we can measure the incident power to the absorber from the filter.
The measured power is lower than 1~\% of total power from the Zotefoam emission. 

For comparison, we also measure the temperature of a filter made of PTFE that has the same geometry as the prototype alumina filter. The PTFE filter is 2~mm thick. The temperature difference with the PTFE filter is $\sim95~\mathrm{K}$ which is significantly larger than that of the alumina filter.

\section{Discussion \label{Disc}} 
We address the scalability in diameter based on our results and discuss the cryogenic compatibility.
Some of the next generation CMB experiments need a window diameter of 500~mm or larger. In this discussion we select PB-2 as an example of such experiments. 
The cooling power of the PB-2 receiver and expected optical heat load are summarized in Table~\ref{Cooling_power}~\cite{Tomaru:2012}. 

\begin{table}
\caption{Summary of cooling power and expected heat load of the PB-2 receiver system as an example of next-generation CMB experiments. PB-2 employs the two pulse tube coolers and provides the 4~K and 50~K stages. One of the pulse tube cooler is dedicated to cool the optical elements and its cooling power for 4~K and 50~K are 0.5~W and 17~W, respectively.  \label{Cooling_power}}
 \begin{tabular}{c|c|c} \hline
     & Cooling power & Expected optical heat load \\ 
    &[W]& [$\mathrm{W}$] \\ \hline
    300~K - 50~K &30 $\times$ 2 &  17~  \\
    50~K - 4~K & 1.4  $\times$ 2 & 0.5~  \\ \hline
  \end{tabular}
\end{table}

PB-2 requires the filter size of 500~mm in diameter that is mounted at the 50~K stage.  This filter absorb the emission from the window. The PB-2 window is assumed to be made of Zotefoam, which absorbs the IR radiation efficiently. Thus, the dominant heat source to the 50~K and 4~K stages is the thermal emission from the room temperature window that is radiatively cooled to around 200~K at the inner surface. The estimated power from the window is about 17~W. We need to remove this heat at the 50~K stage before it reaches the 4~K stage by using the cooling power. 

The temperature of the filter with a given diameter is determined by the thermal conductivity and the thickness, i.e. $\kappa t$ in Equation~\ref{eq:deltaT}, and emissivity.
Although the thicker filter conducts the heat away from the filter more efficiently, the absorption of the incident radiation at the detection band increases with a thicker filter. Therefore, thickness is not a free parameter we can choose for a better thermal performance. When the thickness of the filter is determined, the temperature of the filter equilibrates  based on the conductance and the emissivity of the filter. When the temperature of the filter is significantly higher than the edge temperature, the filter itself becomes an emissive element that contributes significantly to the 4~K stage. 

\begin{figure}
\includegraphics[width=9.5cm]{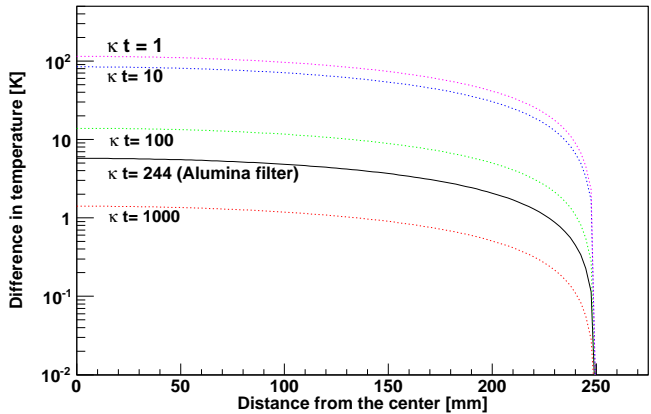}
\caption{ The temperature difference of the alumina filter is shown as a function of the radial distance of the filter. In our model, we assume that the edge temperature is 50~K, the temperature of a Zotefoam window is 200~K, the emissivity of the filter is 1, and the emission from the filter is uniform. Each curve corresponds to $\kappa t =  1, 10, 100, 1000$ in the unit of [mW/K]. A 2~mm-thick alumina and 20~mm-thick PTFE yield $\kappa t =244~\mathrm{mW/K}$ and $2~\mathrm{mW/K}$, respectively. \label{fig:thermal_all}}
\end{figure}

We construct the thermal model based on the thickness and thermal conductivity of the filter in Table~\ref{fig:thermal_conductivity} and \ref{material}.
The incident power from the window is absorbed by the alumina filter, whose power conducts to filter edge and re-emits from the filter surface, respectively.
The emissivity of the filter is assumed as unity in the IR band. Figure~\ref{fig:thermal_all} shows the radial temperature profile of the filter for various materials, including alumina, when the filter diameter is 500~mm and the edge temperature is anchored at 50~K. We compute for various values of $\kappa t$. The IR filter using alumina with the thickness of 2~mm corresponds to $\kappa t=244~\mathrm{mW/K}$. When the alumina filter is used for PB-2, the expected excess temperature at the center of the filter is less than 6~K from the edge temperature.  By contrast, the temperature difference of conventional filters such as polyethylene, Nylon, and PTFE is $\sim120~\mathrm{K}$ and the corresponding re-emission from the filter itself is about 2~W that is significant amount to the available cooling power at the 4~K stage. On the other hand, the expected transmitted radiative power and the re-emitted power from the alumina filter are 20~mW and 45~mW, respectively. These are small enough compared to the cooling power at the 4~K. 

\begin{figure}
\includegraphics[width=9.5cm]{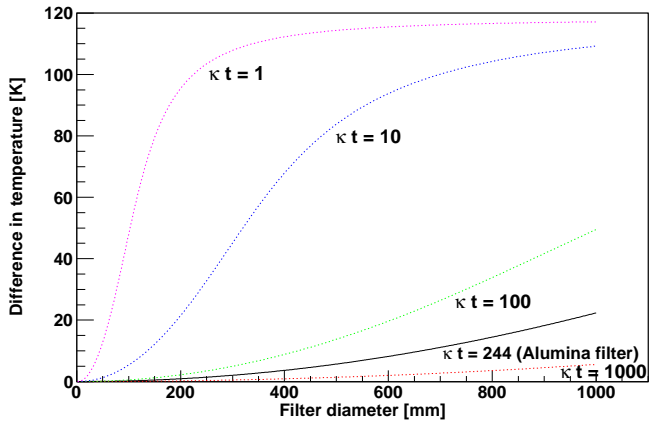}
 \caption{The temperature of the alumina filter is shown as a function of the filter diameter. We assume that the edge temperature is 50~K, and temperature of a Zotefoam window is 200~K.\label{fig:thermal_all_def}}
\end{figure}

We also explore the filter temperature as a function of the filter diameter up to 1000~mm as shown in Fig.~\ref{fig:thermal_all_def}.
The lower the value of $\kappa t$ is, the steeper the temperature rise is due to the limited thermal conductance of the filter. For the case of low $\kappa t$, it is seen that the temperature approaches asymptotically to the value determined purely by the radiative heat exchange. 
The IR filter made of alumina is on the other hand still effective for a diameter beyond 500~mm.


 \section{Conclusion}
We have developed a novel IR filter using alumina and characterized its thermal and optical performances. It satisfies the requirements of the next-generation CMB polarization experiments. We also show the extendability for a future large-aperture millimeter-wave telescope with a very large filter diameter up to 1~m. The IR filter using alumina can be used not only for a telescope application but for any millimeter wavelengths experiment that needs high transmission at millimeter wave while blocking the IR radiation for thermal compatibility with a cryogenic system.

\begin{acknowledgments}
We express our gratitude to Prof. Hiroshi Matsuo and Tom Nitta for providing us with useful information on the measurement of optical properties using the FTS at NAOJ.  We would like to thank Osamu Tajima for the measurements of filter temperature presented in this study.
The following people are acknowledged for their assistance and comments in the filter development and characterization:  Masaya Hasegawa, Suguru Takada, Hideki Morii, Michael Myers, Kam Arnold, Zigmund Kermish, Iwao Murakami and the PB-2 collaboration.
This work was supported by MEXT KAKENHI Grant Numbers 21111002, 24111715,
and JSPS KAKENHI Grant Numbers $25\cdot3626 $, 24740182.
\end{acknowledgments}

\end{document}